\documentclass[twocolumn,showpacs,preprintnumbers,amsmath,amssymb,superscriptaddress]{revtex4}
\usepackage{graphicx}
\usepackage{dcolumn}
\usepackage{bm}
\usepackage{natbib}
\begin{document}
\title{Intermediate Low Spin States in a Few-electron Quantum Dot in the $\nu\le 1$ Regime}
\author{Y. Nishi}

\affiliation{Department of Physics, University of Tokyo, Hongo, Bunkyo-ku, Tokyo 113-0033, Japan}

\author{P.A. Maksym}

\affiliation{Department of Physics and Astronomy, University of Leicester, Leicester LE1 7RH, UK}

\author{D.G. Austing}

\affiliation{Institute for Microstructural Sciences M23A, National Research Council, Montreal Road, Ottawa, Ontario K1A 0R6, Canada}

\author{T. Hatano}

\affiliation{ICORP Spin Information Project, JST, Atsugi-shi, Kanagawa 243-0198, Japan}

\author{L.P. Kouwenhoven}

\affiliation{Department of Applied Physics, Delft University of Technology, P.O.Box 5046, 2600 GA Delft, The Netherlands}

\author{H. Aoki}

\affiliation{Department of Physics, University of Tokyo, Hongo, Bunkyo-ku, Tokyo 113-0033, Japan}

\author{S. Tarucha}

\email{tarucha@ap.t.u-tokyo.ac.jp}

\affiliation{ICORP Spin Information Project, JST, Atsugi-shi, Kanagawa 243-0198, Japan}

\affiliation{Department of Applied Physics, University of Tokyo, Hongo, Bunkyo-ku, Tokyo 113-8656, Japan}

\date{\today}

\begin{abstract}

We study the effects of electron-electron interactions in a circular few-electron vertical quantum dot in such a strong magnetic field that the filling factor $\nu\le 1$. We measure excitation spectra and find ground state transitions beyond the maximum density droplet ($\nu=1$) region. We compare the observed spectra with those calculated by exact diagonalization to identify the ground state quantum numbers, and find that intermediate low-spin states occur between adjacent spin-polarized magic number states.

\end{abstract}


\maketitle

Few-electron quantum dots, or ``artificial atoms", exhibit intriguing effects of quantum mechanical confinement and electron-electron interactions. Both of these effects can be significantly modified by a magnetic field. For example, application of high magnetic fields strengthens electron correlation in quantum dots because the single-particle states become nearly degenerate to form Landau levels, and all the electrons occupy the lowest Landau level when the magnetic field is strong enough~\cite{PRL71_613,PRL77_3613,PRL88_256804}. As the magnetic field is increased from zero, several ground state transitions to higher orbital- and spin-angular momentum states are induced because of Coulomb repulsive interactions, exchange interactions~\cite{PRB45_11419,Nature379_413,PRL74_785,PRB59_2801,PRL83_3270,PRL84_2485,JJAP38_372,PRB72_085331} and correlation effects~\cite{PRL93_206806}. When the field reaches the point where the filling factor $\nu=1$, the ground state becomes a spin-polarized maximum density droplet (MDD) \cite{Nature379_413,PRL74_785,AJP46_345,Oosterkamp}. This state has $N$ electrons occupying the lowest $N$ angular-momentum states in the lowest Landau level and it persists over a wide range of magnetic fields because of the favourable balance of Coulomb repulsive and exchange forces. Further increase of the magnetic field and the resultant wavefunction shrinkage increase the importance of electron-electron interactions and the MDD eventually gives way to a sequence of strongly correlated states which occur in the $\nu\le 1$ regime~\cite{Oosterkamp,PRL65_108}. This is very similar to the fractional quantum Hall effect where electron correlation plays an important role in determining the ground states.

In this strong-correlation regime beyond the MDD, electrons in a quantum dot are predicted to form an ``electron molecule", a state in which electrons rotate and vibrate around a specific classical equilibrium configuration~\cite{PRB53_10871}. To satisfy the Pauli exclusion principle, the ground state angular momentum ($L$) of an electron molecule must belong to a specific series of values called ``magic numbers" and these values depend on both the total spin ($S$) and the rotational symmetry of the molecule~\cite{PRB53_10871,PRB51_7942,PRB52_2978,JPSJ65_3945,PhysicaB249_214}. For $N\le4$ only one series of magic numbers occurs (with interval $\Delta L$=$N$) but for $N\ge5$ multiple series of magic numbers can occur because more than one type of equilibrium configuration is possible~\cite{PRB53_10871,JPSJ65_3945,PhysicaB249_214} and this can open a way to explore novel physics associated with mixing of molecular configurations. Thus quantum dots in the $\nu\le 1$ regime are extremely interesting for the study of strong correlation.

However, the few existing experimental studies of quantum dots in this regime~\cite{Nature379_413,PRL74_785,Oosterkamp} are mainly concerned with the MDD region at large $N$. For example, in \cite{Oosterkamp} the general trends of the ground state evolution for large $N$ are discussed in terms of charge redistribution but the small $N$ regime is not studied. Despite its importance as a model interacting system, a dot in the small $N$, large $B$ regime is poorly understood experimentally and even its ground state quantum numbers have not yet been identified.

In this Letter we report on ground state transitions in the electronic energy spectra in the $\nu\le 1$ regime and, to identify the ground state configurations, we compare the observed spectra with exact calculations. To do this we have developed a new theoretical model of a pillar dot, whose parameters are derived from experiment, so that data for individual dots can be analysed \cite{maksym}. We focus on $N=5$ and $N=3$ quantum dots, since the $N=5$ dot is the smallest system with two types (pentagonal and square+center) of molecular configuration (hence two series of magic numbers), while the $N=3$ dot has the simplest non-trivial molecular configuration (triangular).

\begin{figure}
\begin{center}
\includegraphics*[width=70mm, height=60mm]{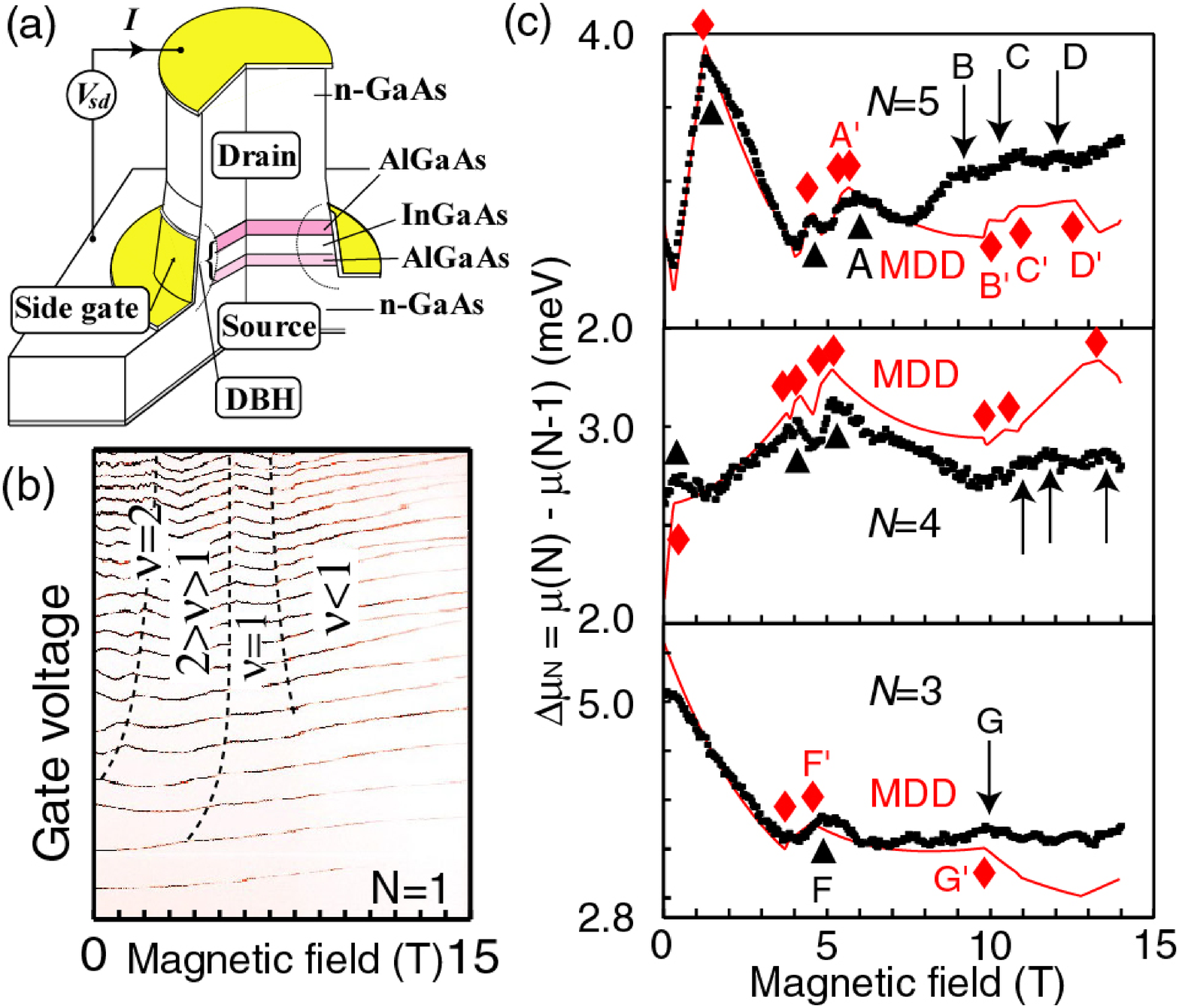}
\end{center}
\caption{(a) Schematic diagram of quantum dot mesa. (b) $B$ dependence of the Coulomb oscillation peaks. (c) Experimental and theoretical $\Delta \mu_N$ against $B$ for $N=3$, 4 and 5.}
\label{fig1}
\end{figure}

Our sample is a circular vertical quantum dot mesa of diameter 520nm (Fig.~\ref{fig1}(a)). Details of the layer structure and device mesa can be found in Ref.~\cite{PRL77_3613}. Unlike for a lateral quantum dot geometry, tunneling rates between the dot and contacts are not drastically suppressed even in high magnetic fields (up to 15~T) because the magnetic field is applied parallel to the current in our device geometry. Our measurements are performed in a dilution refrigerator and the electron temperature is estimated to be below 100mK.

The current $I$ flowing vertically through the dot is measured as a function of the gate voltage $V_g$ at a fixed dc source-drain voltage $V_{sd}$. For small $V_{sd}$ ($<100\mu$V), current flows only through the ground states. Then the position of the $N$th current (Coulomb oscillation) peak $V_{g}(N)$ is a measure of the $N$th ground state electro-chemical potential $\mu(N)$. For finite $V_{sd}$, on the other hand, the excited states in the transport window as well as the ground state contribute to the current. We can then obtain excitation spectra in which the excited states appear as step-like features within a broad current stripe~\cite{Science278_1788}.

The magnetic field ($B$) evolution of the Coulomb oscillation peaks for $N=1\sim 20$ is shown in Fig.~\ref{fig1}(b). This result is similar to that reported earlier~\cite{Oosterkamp}. As discussed in refs.~\cite{PRL77_3613,Oosterkamp}, atomic features such as a shell structure and deviations due to Hund's first rule indicate that our dot has high rotational symmetry. Regions of $\nu>2$, $2>\nu>1$, $\nu=1$ and $\nu<1$ can be distinguished by the overall behavior of the peaks~\cite{Oosterkamp}. The peak spacing, $\Delta V_{g}=V_{g}(N)-V_{g}(N-1)$, is not strongly $N$-dependent in the MDD region. This is no longer the case when the $\nu<1$ region is entered, which implies that a new energetic structure is induced by correlation. Now we focus on the evolution of $\Delta V_{g}$ to study the ground state transitions, which has not been done before for $\nu<1$.

The peak spacings are related to the $N$-electron ground state energy $E_N$ via the ``addition energy" $\Delta \mu_N=\mu(N)-\mu(N-1) = E_{N} - 2 E_{N-1} + E_{N-2}= e(\alpha_N(V_g,B) V_g(N) - \alpha_{N-1}(V_g,B) V_g(N-1)) \sim e\bar{\alpha}\Delta V_g$ where $\alpha$ is an electrostatic leverage factor~\cite{kouwenhoven01} and $\bar{\alpha} = (\alpha_N + \alpha_{N-1})/2$. We use experimental values of $\bar{\alpha}$ to convert $\Delta V_g$ to $\Delta \mu_N$ and compare the results with $\Delta \mu_N$ calculated from $E_N$ values obtained by exact diagonalization. The effects of electron-electron interactions, screening, finite dot thickness, image charges, and the $V_{g}$ dependence of the confinement potential are incorporated realistically. The parameters in the theoretical model, the confinement energy, $\hbar \omega = 4.80 + 0.021 (N-1)$ meV and screening length $\lambda = 15.3$ nm, are found by fitting data up to 10~T~\cite{maksym} and the $g$-factor, $|g^*| = 0.3$ is estimated from the Zeeman splitting observed at high fields.

Figure~\ref{fig1}(c) shows experimental (black) and theoretical (red) $\Delta \mu_N$ values for $N=3, 4$ and 5. There is generally excellent agreement especially for $B<10$~T where the RMS deviation is less than $\sim0.15$meV. For $B<8$~T (i.e. $\nu\ge1$) almost all the ground state transitions predicted in the theory (indicated by diamonds) are observed as upward kinks (triangles) in the experimental results.  The MDD region extends over a wide magnetic field range between two kinks in each curve. Some of the predicted transitions are missing probably due to the resolution of our experiments. For $B>8$~T there are discrepancies in the magnitude of the experimental and theoretical values of $\Delta \mu_N$ but the positions of features agree well in both Fig.~\ref{fig1}(c) and excitation spectra (discussed below). To investigate the cause of the magnitude discrepancie we used the field range up to 14~T (instead of 10~T) to fit the model parameters and found that the positions of the features did not agree in either the low or high field regimes. This suggests that the magnitude discrepancies are caused by imprecise knowledge of $\bar{\alpha}$ in the high field regime~\cite{disorder}. However precise knowledge of $\bar{\alpha}$ is {\it not} needed to interpret the positions of features in the excitation spectra so we can apply our model to identify ground state quantum numbers.

\begin{figure}
\begin{center}
\includegraphics*[width=70mm, height=60mm]{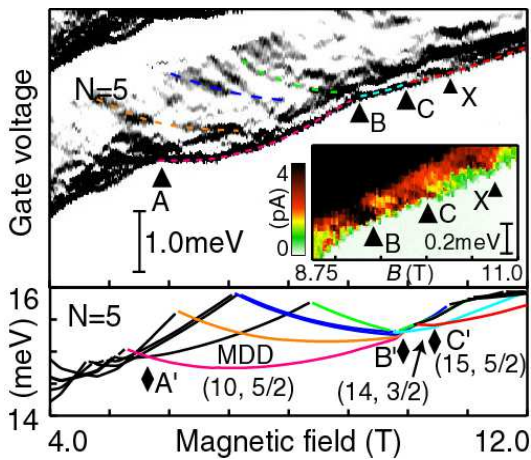}
\end{center}
\caption{Top: Measured excitation spectrum for $N=5$ for $B=4 - 12$~T with $V_{sd}$=1.8mV: $dI/dV_{g}$ is plotted to show sharp changes. (Inset: magnitude of $I$ boundaries between different colors indicate excited states). Bottom: Calculated energy spectrum. Colored lines are guides to the eye.}
\label{fig3}
\end{figure}

We begin with $N=5$. The observed excitation spectrum for $B=4$ - 12~T is shown in the upper panel of Fig.~\ref{fig3}, where regions of positive (zero and negative) $dI/dV_{g}$ appear black (white) in the $B$-$V_{g}$ plane. The lower edge of the stripe delineates the ground state.  Since we are interested in the ground state transitions, we focus on the relative positions of the ground state and a few of the lower excited states. We find the MDD ground state (pink broken line) between two kinks of the lower edge of the stripe at $B=5.9$~T (labelled A) and 9.3~T (B), as already seen in Fig.~\ref{fig1}(c). The area around B is magnified in the inset, where the magnitude of $I$ is color coded~\cite{Intensity_plot}. As we go above the lower edge of the stripe the color abruptly changes from green to red or black for $B\le 9.3$~T and $B\ge 10.0$~T, which indicates an onset where an excited state can contribute to the current. Indeed, we can see that an excited state for $B\le 9.3$~T becomes the ground state between 9.3~T (B) and 10.0~T (C), which then becomes an excited state again for $B>10.0$~T. Namely, an intermediate ground state emerges at the collapse of the MDD state. The transitions B and C agree well with those in the calculated spectrum (defined as $E^*_N -E_{N-1} - \mu_c$ where $E^*_N$ is the $N$-electron excited state energy and $\mu_c$ is an approximation to the contact electro-chemical potential \cite{maksym}) which is shown in the lower panel of Fig.~\ref{fig3}.  This shows that a partly spin-polarized ground state with $(L, S)=(14, 3/2)$ appears between the fully spin-polarized magic-number states $(10, 5/2)$ (between A$^\prime$ and B$^\prime$) and $(15, 5/2)$ (beyond C$^\prime$). The calculation also reproduces other important features in the experimental spectrum. For example, parts of the excited states along the orange, blue and green broken lines in the $dI/dV_{g}$ plot of Fig.~\ref{fig3} are reproduced by the calculation.

\begin{figure}
\begin{center}
\includegraphics*[width=70mm, height=80mm]{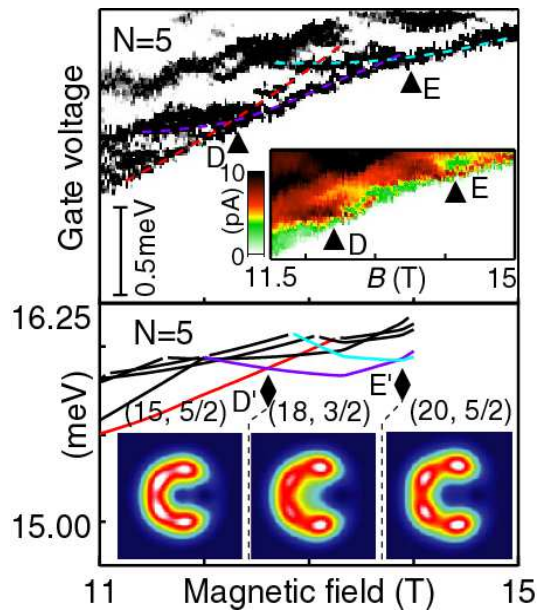}
\end{center}
\caption{Measured excitation spectrum for $N=5$ for $B=$11 - 15~T (top), and the calculated spectrum (bottom). Colored lines in both spectra are guides to the eye. Upper inset: magnitude of $I$. Lower inset: pair-correlation functions.}
\label{fig4}
\end{figure}

\begin{figure}
\begin{center}
\includegraphics*[width=70mm, height=45mm]{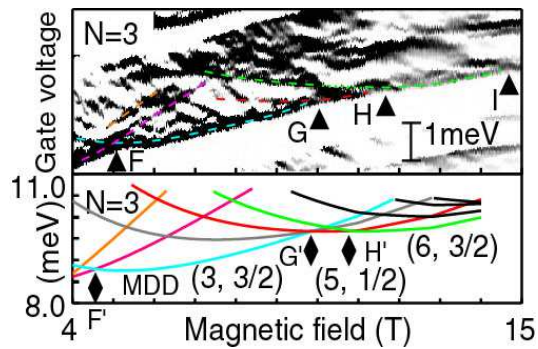}
\end{center}
\caption{Top: Measured excitation spectrum for $N=3$ for $B=4$ - 15~T with $V_{sd}=$2.7mV. Bottom: Calculated energy spectrum. Colored lines in both spectra are guides to the eye.}
\label{fig5}
\end{figure}

Within the stripe there are several grey ``patches". Those near the lower edge of the stripe can be attributed to ``emitter states", which result from density of state fluctuations in the heavily doped contacts~\cite{PRL78_1544}. Emitter state features are distinguished from excited state features because (i) they do not smoothly connect to the ground state and/or (ii) similar patterns can be found in neighboring stripes (i.e. they are $N$-independent) and/or (iii) they appear as fluctuations or peak-like features in the $I-V_{g}$ characteristics, whereas excited states appear as step-like features (See the insets to Figs. 2 and 3)~\cite{Intensity_plot}. For example, a transition-like structure observed around the point labelled X in Fig.~\ref{fig3} can be attributed to the emitter states because (i) it ``touches" the ground state so steeply that it cannot smoothly connect to the ground state, and (ii) similar features are found for $N=4$ and 6 at the same magnetic field. The ``patches" not very close to the lower edge can arise both from emitter states or higher lying excited states which are so dense that it is almost impossible to identify them experimentally. These general statements also apply to Figs.~\ref{fig4} and \ref{fig5}.

No ground state transitions occur for $B=10.0$ - 12.0~T in Fig.~\ref{fig3}, so we now explore the higher magnetic-field region, B$>$12~T for $N=5$. Fig.~\ref{fig4} shows the excitation spectrum for $B=11$ - 15~T, where the upper panel and inset, respectively, show $dI/dV_{g}$ and $I$. We can immediately see that the ground state for $B\le12.4$~T (red broken line) gives way to a new ground state (violet broken line) at $B=12.4$~T (D), which is in turn taken over by another excited state (light blue broken line) that becomes the next ground state at $B=14.0$~T (E). These transitions are also identified as color changes in the $I$-intensity plot inset. If we turn to the calculation in the lower panel, we can identify the ground state between D and E as the partly spin-polarized $(L,S)=(18, 3/2)$ state between D$^\prime$ and E$^\prime$, and the ground state for $B\ge 14.0$~T as the spin-polarized magic-number state $(20, 5/2)$ to the right of E$^\prime$.

Let us now compare the above with the $N=3$ excitation spectrum (Fig.~\ref{fig5}, upper panel). A down-going excited state (light blue broken line) changes places with the up-going ground state (pink) at $B=5.0$~T (F). This indicates a transition to the MDD state, $(L,S) = (3,3/2)$, for $B= 5.0$ - 10.0~T, consistent with Fig.~\ref{fig1}(c). There are several dark patches (not reproduced by theory) intersecting with the MDD region, but we attribute them to emitter states because they shift too rapidly with magnetic field to connect to the ground state smoothly. However, it can be seen that a down-going excited state (red broken line) crosses the MDD ground state at $B=10.0$~T (G), for which the resulting kink is also observed at $B=10.0$~T in Fig.~\ref{fig1}(c). This marks the collapse of the MDD state, and a transition to a new ground state. We also find another down-going excited state (green broken line). This excited state crosses the ground state (to the right of G) at $B=11.6$~T (H). Hence an intermediate ground state (red broken line) occurs between G and H. Another transition occurs at $B=14.5$~T (I) and one is predicted to occur just above 15~T (not shown). The discrepancy may be due to disorder~\cite{disorder}. If we compare the observed transitions to those in the theoretical spectrum in the lower panel, we can identify the intermediate state as the low spin state $(5, 1/2)$ that appears between two spin-polarized magic-number states $(L,S)=(3,3/2)$ (MDD state) and $(6, 3/2)$ (to the right of H).

To summarize, for $N=5$, we have observed and identified the ground state transition from the MDD state ($(L,S)=(10,5/2)$) to the next spin-polarized magic-number state $(15,5/2)$ via the low-spin state $(14,3/2)$, and a higher transition to another spin-polarized magic number state $(20,5/2)$ via the low-spin state $(18,3/2)$. For $N=3$, the MDD state $(3,3/2)$ makes a transition to the next spin-polarized magic-number state $(6,3/2)$ via the low-spin state $(5,1/2)$.

The nature of the low spin states can be interpreted by using the electron-molecule picture~\cite{PRB53_10871}. Specifically, for $N=5$ and $S=5/2$ the magic numbers $L=14$ and 18 correspond to a `square + center' configuration whilst the pentagonal configuration is forbidden. In contrast, when $S=3/2$, electron-molecule theory predicts that {\em both} pentagonal {\em and} square+center configurations are possible for $L=14$ and $18$. This has interesting consequences which we investigate by calculating ground state pair-correlation functions. The pair correlation functions (Fig.~\ref{fig4}) for the spin-polarized molecular states $(15, 5/2)$ and $(20, 5/2)$ show pentagonal symmetry but the one for the $(18, 3/2)$ intermediate state has a plateau at the center instead of a valley. This suggests that the $(18, 3/2)$ state is well approximated by a mixture of molecular states with pentagonal and `square + center' symmetry.

In conclusion, we have measured energy spectra of $N=3$ and $5$ quantum dots and observed various ground state transitions. We have compared them with theoretical calculations to assign ground state quantum numbers. We have identified intermediate low-spin states in the $\nu\le 1$ region located between spin-polarized magic number states and explained their origin using the electron-molecule picture.


DGA, LPK, and ST are partly supported by DARPA-QUIST program (DAAD 19-01-1-0659). ST is grateful for financial support from the Grant-in-Aid for Scientific Research A (No. 40302799), SORST-JST and IT program MEXT. PAM thanks the University of Leicester for the provision of study leave. We are pleased to thank H. Imamura for fruitful comments and discussions.


\bibliography{apssamp}

\begin{thebibliography}{99}
\bibitem{disorder}



There may be a small effect from disorder as the length scales of dot states and potential fluctuations in our sample become similar at a field of about 15 T.
\bibitem{Intensity_plot}





A $dI/dV_{g}$ plot can show abrupt changes in $I$ clearly over a wide $B$-range, but it also emphasizes $I$ fluctuations caused by emitter states. In a color-coded $I$ plot it is difficult to show the spectrum over a wide $B$-range because the absolute value of $I$ can change drastically with $B$. However, step-like features due to the excited states can be highlighted in a color-coded plot over a small $B$ range.
\end{thebibliography}

\bibliographystyle{apsrev}



\end{document}